# Tunable Multifocal THz Metalens Based on Metal-Insulator Transition of VO$_2$ Film

Roya Kargar [1], Kasra Rouhi [1], and Ali Abdolali [1*]

[1]Applied Electromagnetic Laboratory, School of Electrical Engineering, Iran University of Science and Technology, Tehran, 1684613114, Iran

*Corresponding author. E-mail: abdolali@iust.ac.ir (Ali Abdolali)

*Abstract* – **Recently, metalenses which consist of metasurface arrays, have attracted attention due to their more condensed size in comparison with conventional lenses. In this paper, we propose a reconfigurable coding metasurface hybridized with vanadium dioxide (VO2) for wavefront manipulation at terahertz (THz) frequencies. At room temperature, the unit-cell can reflect as a "1" bit under linearly y polarized illuminated waves. Besides, when the temperature is increased, VO2 would be in a fully metallic state; therefore, unit-cell can act as a "0" reflection phase. Furthermore, by changing the unit-cells arrangements on a metalens surface, the proposed device can focus the incident beam at any position according to a particular design. Numerical simulations demonstrate that the designed VO2-assisted metasurface can generate one and multi-focal spot in reflection mode as expected. Also, theoretical results depict an excellent agreement with obtained simulation results. The presented metalens has notable potential in THz high-resolution imaging and optical coding.**

I. Introduction

Metasurfaces, the two-dimensional version of metamaterials, have recently attracted significant consideration due to their ability to manipulate arbitrary electromagnetic wavefronts by adding extraordinary field discontinuities over the interface [1,2]. Planar equivalents of conventional bulky optical devices such as lenses [3,4], holograms [5,6], absorbers [7,8], optical signal processors [9], beam deflectors [10–12], polarizers [13], and so on have been experimentally demonstrated in different electromagnetic frequencies. These devices consist of a subwavelength element arranged on a two-dimensional surface with a specific order. Indeed metasurfaces are mimicking the phase profile of the typical bulk optical devices. Such metasurface-based flat devices represent a new class of optical and THz components that are compact and lightweight. Although metasurface application and fabrication methods improved in recent years but reconfigurable metasurface devices are a significant missing link in modern pragmatic devices and systems.

Mostly metasurfaces are composed of passive building blocks that cause a lack of tunability and reconfigurability of response. Therefore, one metasurface-based device only grants one or a few functions that are inherently limiting their practical impact. Among these devices, compact lenses with reconfigurable focal points have various applications. Hence, various methods have been developed to make such devices. Deformable solid and liquid-filled lenses with mechanical [14], electrowetting [15,16], electromechanical [17], and thermal tuning [18] mechanisms have been proposed in several pieces of research. These devices are more compact than conventional varifocal lenses and have different tuning rates ranging from few Hz to a few hundreds of Hz. On the contrary, liquid crystal lenses with a tunable focus spot have higher tuning speeds, but they suffer from polarization dependence and limited tuning range [19,20]. Freeform optical elements that can tune the focus distance upon lateral displacement of the components have also been expressed [21]. These devices are based on the mechanical movement of massive elements and are consequently not very compact nor fast. Highly tunable elastic dielectric metasurface lenses based on stretchable substrates have also been proved, but they have low speeds and require a radial stretching mechanism that might increase the device size [22,23]. Also, controlling the axial movement or angular orientation of the metalens via integrating with the microelectromechanical systems [24,25] and laterally actuating two separate cubic metasurfaces based on the Alvarez lens design [26] are novel approaches for tunable metalens. On the other side, procedures of including tunable materials into metasurfaces for changing the functionality, such as the use of liquid crystals [27], phase-change materials [28,29], graphene [30–32] or others are widespread in various devices. Dynamic responses of these materials enable active THz materials that are excited by external stimuli via photoexcitation, electric or magnetic bias and temperature.

Vanadium dioxide is an advantageous material to accomplish a switchable performance in electromagnetic devices. VO2 is praised for its phase transition feature, which changes the material from a semiconductor to a metallic state at a critical temperature $T_c = 68°C$. This unique characteristic is due to physical and structural variations, which make changes in the material electromagnetic properties [33]. Atomic-level deformation in phase change materials has capable of offering drastic variations in material properties over a broad spectral range during the phase transition. These features make VO2 a promising phase-change material to be utilized in tunable metasurfaces. By designing sub-wavelength structures, VO2 has been employed to fabricate tunable meta-devices working at THz and other optical frequencies. The dielectric properties of VO2 could be impressed by thermal, electrical, optical, and mechanical stimuli, and the substantial point in this process is reversible and has a hysteresis associated characteristic [34]. Recently VO2-assisted metasurfaces attract scientist's attention and utilized in a diverse application for electromagnetic wave manipulation [35–37].

In this manuscript, we propose a high-performance, one-bit tunable metasurface by utilizing VO2 film for the THz applications. According to the theory of fuzzy quantification, two dynamically switchable metasurfaces whose reflection phases can possess both of 0 and $\pi$ values, are utilized as "0" and "1" building blocks of 1-bit coding metasurface. By adjusting the vanadium dioxide temperature in each unit cell and implementing the required reflection profile for wave concentration, an exact focus point is obtained at an arbitrary focal point. The results illustrate that the proposed THz metalens exhibits precise focusing at a frequency of 1.6 THz, and it can include the whole half-space. Also, this structure has a valuable capability to concentrate reflected waves in more than one focus region. In the following subsections, we will describe our proposed one-bit digital metasurface and provide the required mathematical calculations and analysis to utilize VO2-assisted metasurface as multifocal metalens. We have investigated all simulation results by full-wave simulation software and compared them with a theoretical approach.

## II.     Vanadium Dioxide

Vanadium dioxide is known for exhibiting a reversible first-order phase transition from a monoclinic to a tetragonal crystalline structure at a transition temperature of 68 ℃. Owing to its association with a strong variation of both the resistivity and the optical dielectric constant, this solid phase transition attracts substantial interest. VO2 is transparent at infrared frequency range at the insulator phase, while VO2 at the metallic phase is opaque at the most spectral frequencies. Due to its low photon energy, THz radiation is sensitive to the change of free carrier density in a phase transition [38].

Another VO2 property is its sub-picosecond time duration over which the solid phase transition occurs. All these characteristics make VO2 extremely interesting not only for the fundamental understanding of the physics of this semiconductor to metal transition but also for near room temperature switching devices. As mentioned in other references, the electrical resistance of the VO2 films is a function of temperature for both heating and cooling cycles. This cycle is not unique for the heating and cooling processes in transition temperature. The metal-insulator transition behavior of VO2 films makes it as two dissimilar materials in which one of them is an insulator, and the other one is metal. Its relative permittivity is about 9 in the insulating state, while the conductivity is smaller than 200 S/m. When VO2 passes the transition temperature, it comes to the metallic state. In this case, relative permittivity is constant, but the conductivity increases as high as an order of $10^5$ [39].

### III. Metasurface Structure

Hereby we introduce VO2 in a suitably configured structure such that its reflection phase is either 0° or 180° depending on the state of VO2. Based on arrays of such a metasurface, we demonstrate dynamic electric field distribution at the operating frequency of 1.6 THz from rigorous full-wave simulations. The proposed metasurface consists of three layers of gold, silicon, and combination of VO2 and gold, respectively, from bottom to top. Each unit cell consists of three gold fragments with a thickness of $t_m$=200 nm. Also, we have incorporated four similar VO2 fragments into the gold gaps, which can be programmed individually. The spacer dielectric substrate is lossy silicon, which has the relative permittivity of $\varepsilon_r$=11.9 and $\delta$=0.00025. Finally, a gold ground plane has been embedded in the last layer to prevent electromagnetic radiation transmissions. *Figure 1* illustrates a schematic of a designed unit cell and all the required geometry parameters. The reprogrammable device can be employed to deliver the desired bias voltages at its output and control each meta-atom temperature to set the operational status of each coding element, individually. Consequently, by a real-time switch between arbitrary coding patterns, a different focal spot can be achieved.

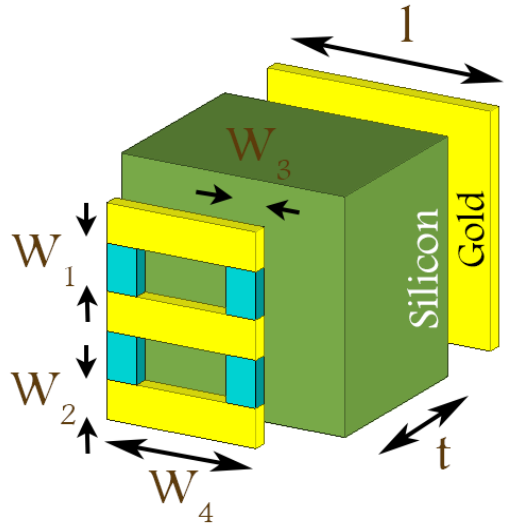

*Figure 1. The schematic of the proposed unit cell and constituent materials have been depicted in this figure. The yellow material is gold, the green material depict lossy silicon, and the blue material is VO2 in both insulator and lossy metal state. In the proposed structure, $w_1$=3 μm, $w_2$=2.5 μm, $w_3$=2 μm, $w_4$=10 μm, $l$=15 μm, $t$=12 μm.*

All the designs are carried out using CST Microwave Studio with the frequency-domain solver to extract the reflection phases and amplitude coefficients for an infinite array of "0" and "1" coding elements. To simulate the suggested structure, periodic boundary conditions are applied in the x and y directions to consider the mutual coupling influence between adjoining elements. Also, the Floquet ports are assigned to the z-direction to illuminate incident y-polarized waves. In metalens design, the conductivity of VO2 in each unit cell is varied to tailor the required phase change, while the construction of the whole metasurface is kept constant. When VO2 is insulating, an element is coded by "1" and correspondingly, an element with VO2 in the metal state is coded by "0".

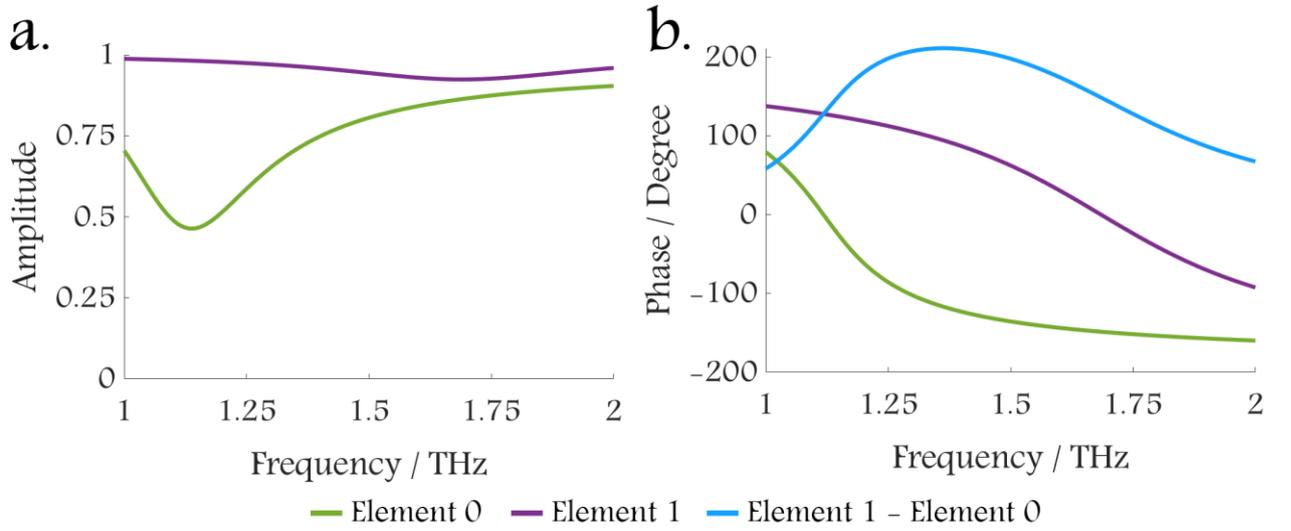

*Figure 2. Reflection coefficient response of the (a) amplitude and (b) phase of reconfigurable digital meta-atom under y-polarization illumination from 1 to 2 THz. According to extracted results from simulation, 1.6 THz selected as a primary working frequency.*

**IV.     The Focusing Properties of Tunable Metalens Structure**

To demonstrate how VO2-assisted unit cells are utilized in metasurface designs for manipulating the scattered field concentration dynamically, we first describe the required phase distribution for beam concentration. The waves reflected from the metasurface must constructively interfere at the focal point similar to the waves that are transmitted from a conventional lens. To construct a planar lens, we spatially distribute the coding unit cell with varying reflection phase to realize the hyperbolic phase profile. The target phase profile of the metalens is given by the following expression [40]:

$$\varphi_{ref}(x,y) = k(\sqrt{(x-x_{foc})^2 + (y-y_{foc})^2 + z_{foc}^2} - z_{foc}) \qquad (1)$$

Where $k=2\pi/\lambda$ is the wave vector in desired frequency, A(x,y,0) is a point on the metasurface, and A'($x_{foc}$,$y_{foc}$,$z_{foc}$) is a predetermined focal point in half-space. In this equation, we assume that the metalens is positioned at the z=0 plane, and the center of each unit cell represents the element location.

## V. Focus Results

In this section, we will investigate our structure capability to concentrate reflected beam in a predetermined focal spot. In the first example, phase-change reconfigurable metasurface designed to concentrate radiated waves at the metasurface normal axis. Indeed we have started to arrange digital building block in origin of coordinates and utilized required parabolic phase distribution. *Figure 3 (a)* shows a metasurface phase profile to concentrate the beam at x=0 μm, y=0 μm, z=170 μm. This structure composed of 80 × 80 elements, which is equal to 6.4λ × 6.4λ at 1.6 THz approximately. Obtained results for electric field distribution have been demonstrated in *Figure 4 (a-c)*. As we can see in these figures, the metasurface produces efficient electric field concentration very close to the desired point. As exhibited in *Figure 4 (a),(b)*, side lobes appear around the focal spot due to the finite and discrete nature of the proposed metasurface. Side lobes level value and focal point position accuracy can be improved with larger metasurfaces size or smaller unit cell dimension. To evaluate simulation results, we have utilized a theoretical approach to calculate the electric field distribution; then, we have compared theoretical and simulation results.

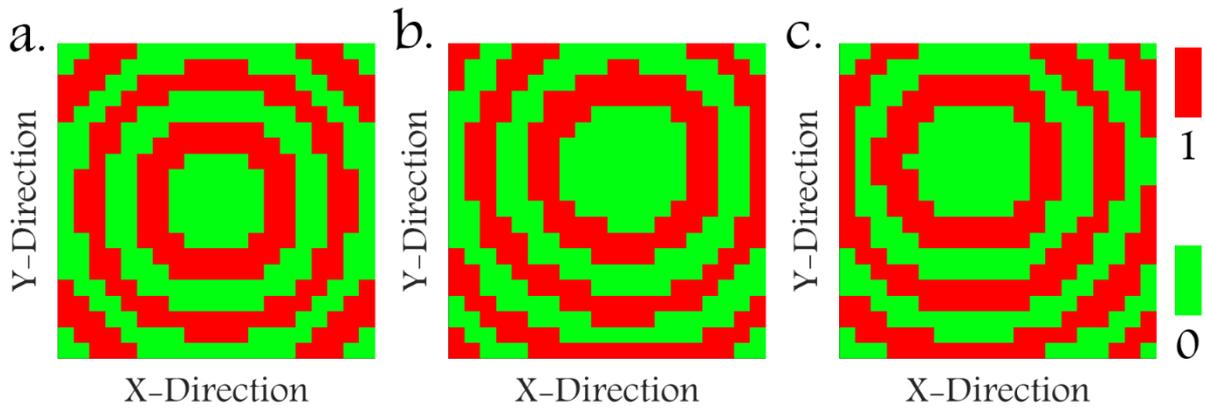

*Figure 3. Encoded unit cells arrangement used to achieve the focal spot at (a) x=0 μm, y =0 μm, z=170 μm (b) x=70 μm, y=140 μm, z=270 μm (c) x=-100 μm, y=155 μm, z=230 μm.*

According to the Huygens' principle, each radiative elements can be assumed as a secondary source, so the meta-atoms are equivalent to a point sources. It gives rise to a diffraction effect at the rear surface of the structure. Due to the particular phase distribution of structure, the effect could be equivalent to the combined effect of a zone plate and an echelle grating [41]. To confirm our analysis, we departed from the Fresnel diffraction method to look through the electric field distribution [6]

$$E(r_1) = \iint E(r_0) \frac{-jz}{\lambda} \frac{\exp(-jkd_{01})}{d_{01}^2} ds \qquad (2)$$

Where $E(r_1)$ demonstrates the electric field at point $r_1$ on the observation point, $E(r_0) = U_{m,n} \exp(j\varphi_{m,n})$ is the electric field at point $r_0$ on the metasurface, whereas this reflection distribution is equivalent to reflection amplitude and phase profile, z is a distance between metasurface and observation point, and $d_{01}$ is the distance between points $r_0$ and $r_1$. Also, according to the digitalized nature of the structure, we can rewrite Fresnel diffraction as below:

$$E(r_1) = \sum_M \sum_N U_{m,n} \exp(j\varphi_{m,n}) \frac{-jz}{\lambda} \frac{\exp(-jkd_{01})}{d_{01}^2} \qquad (3)$$

The calculated result by utilizing the analytical approach is shown by the green dashed lines in *Figures 4 (a-c)*, which agrees well with the solid purple lines obtained by simulation. For assurance of the meta-device dynamic response, two more examples with different focusing spots were presented, which demonstrate that the focal point can be switched in a real-time manner in the whole space. As a second example, we have selected a focal point far from the normal axis. We have arranged a phase profile according to *Figure 3 (b)* to concentrate reflected beam at x=70 μm, y=140 μm, and z=270 μm. The simulation and analytical results have been shown in *Figures 4 (d-f)*. The simulated phase profile exhibits a proposed device capability to concentrate reflected energy in a different spot out of the normal axis. The whole space scanning capability is the most significant feature of reconfigurable metalens for imaging applications, whereas in the conventional imaging systems scanning has been done by using mechanical movements [25]. Indeed in the VO2-assisted metalens, mechanical tunability replaced by temperature tunability via electrical stimulation. Finally, in the last example, we have examined the surface tunability to concentrate reflected beam at x=-100 μm, y=155 μm, and z=230 μm. The required encoded phase profile and one-dimensional electric field distribution exhibited in *Figure 3 (c)* and *Figures 4 (g-i)*, respectively, which we can observe a good agreement between theoretical and simulation results.

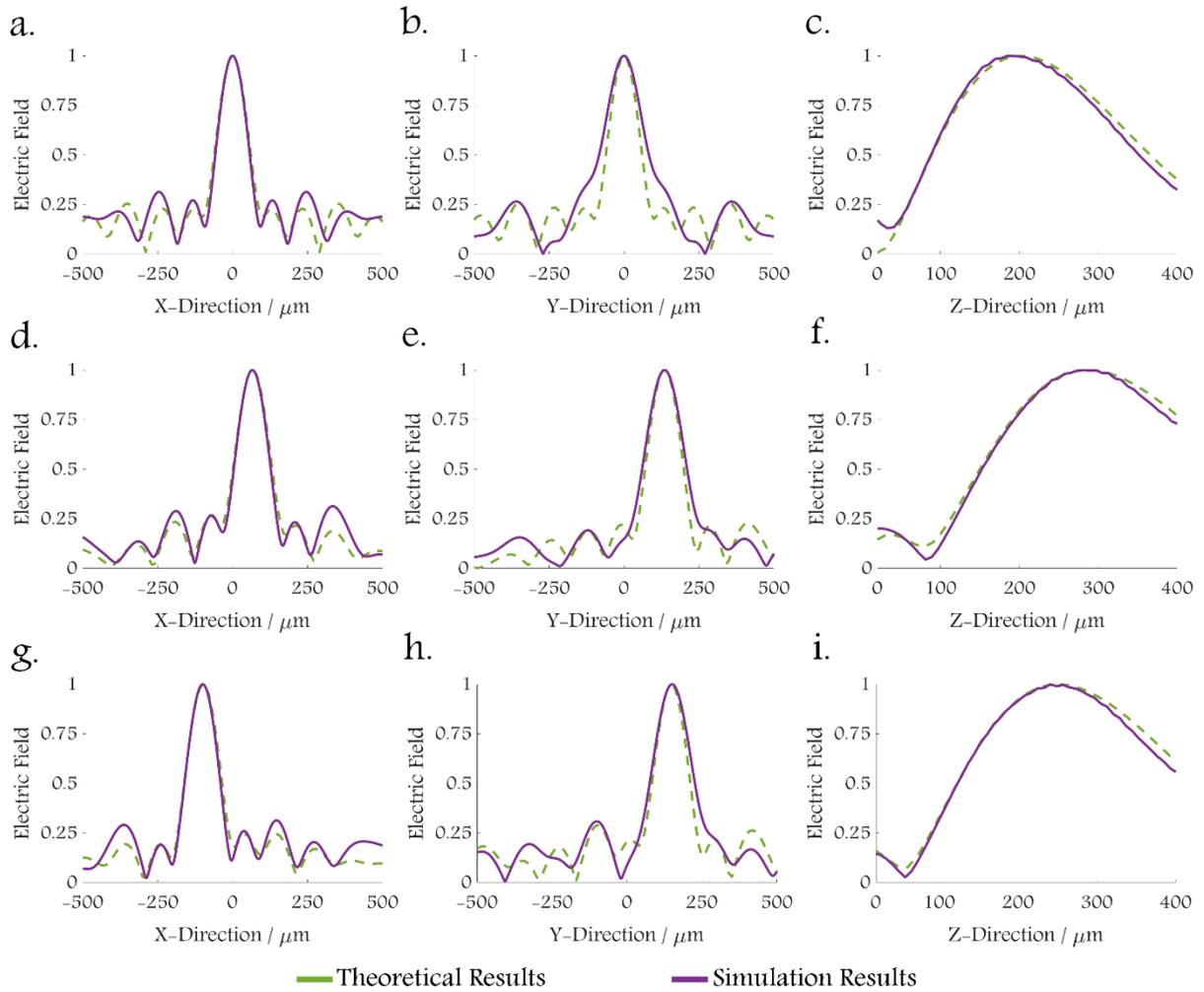

*Figure 4.* One-dimensional electric field distributions of VO2-assisted metalens with the different focusing spot at the working frequency of 1.6 THz in three directions. All figures contain theoretical and simulation results. (a-c) x=0 μm, y=0 μm, z=170 μm (d-f) x=70 μm, y=140 μm, z=270 μm (g-i) x=-100 μm, y=155 μm, z=230 μm.

## VI. Multiple Focus Results

Proposed VO2-assisted metasurface has a valuable feature to concentrate reflected beam at more than one focal spot. To examine this capability, we have divided a structure to multi sections that each section produces one focal point in space. Different distribution of digital elements in each segment leads to multiple focal points only by one planar structure. *Figure 5* illustrates the schematic diagram for a suggested approach to generate two focal spots in different positions, simultaneously.

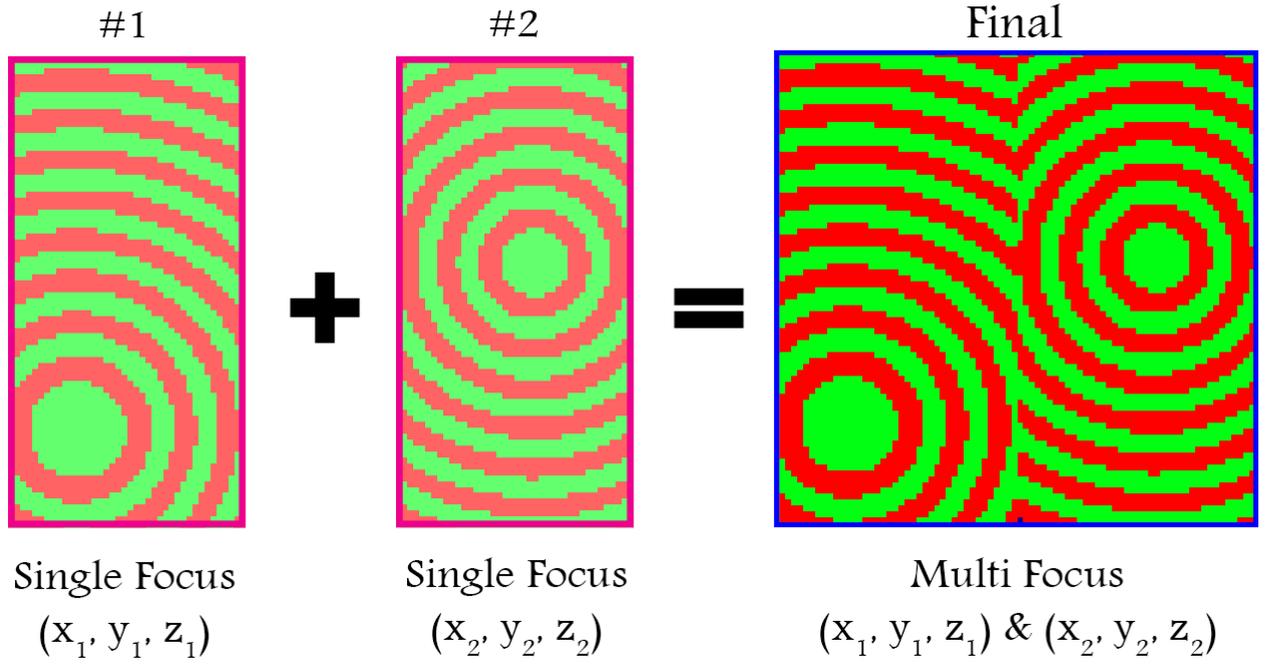

**Figure 5.** *Schematic diagram of multifocal metasurface by using dividing approach.*

For instance, we have calculated required encoded arrangement to concentrate reflected beam at x=-105 μm, y=285 μm, and z=100 μm for the first section and x=185 μm, y=-150 μm, and z=100 μm for the second section. ***Figure 6 (a)*** exhibits simulation result in the focal plane, and ***Figure 6 (b)*** shows the analytical result for this example. In this example, selected focal points are placed in the same plane and have a similar distance with metasurface, but it is possible to concentrate reflected energy in two different planes.

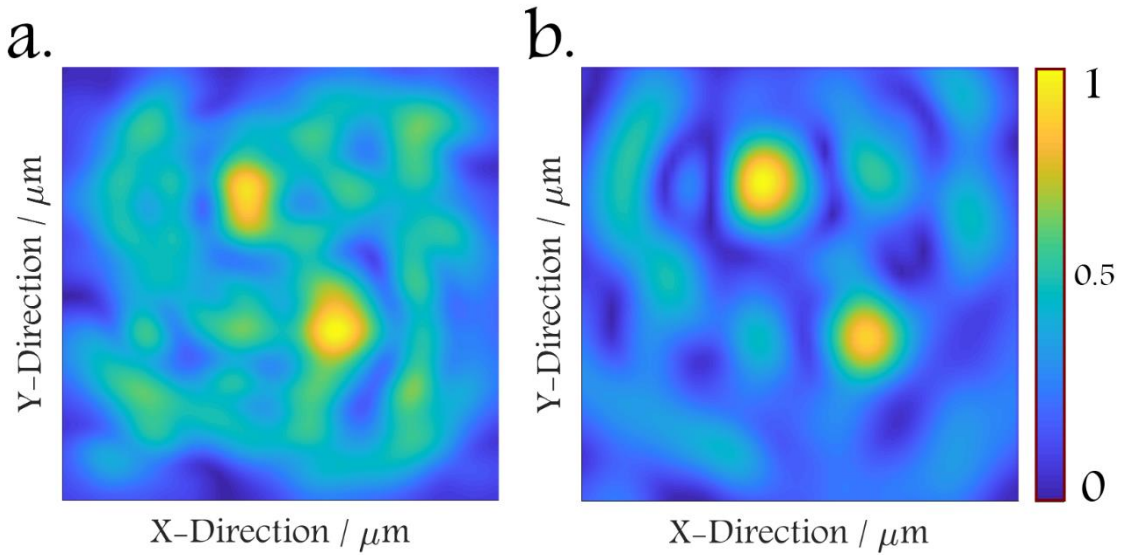

**Figure 6.** *Two-dimensional electric field distribution of VO2-assisted metalens for multifocal application at a frequency of 1.6 THz in similar transverse focal plane z=100 μm. These figures contain obtained results by (a) simulation and (b) theoretical approach.*

## VII. Conclusion

In summary, we have proposed an engineered VO2-assisted metalens for dynamic THz wave concentration. Designed unit cell has a particular capability to tune between two digital states, "0" and "1" by controlling VO2 material temperature. Based on this configuration, a THz metalens with tunable focal point is designed and numerically demonstrated. Besides, we have proposed a dividing method to generate multifocal metalens and examine this method by a numerical and analytical approach. The proposed design and concept have the flexibility to be developed for more than two focal points with high-resolution concentration. The designed VO2-assisted metasurface has excellent potentials for tunable, multifunctional, and integrated THz devices.